\documentclass{moriond}

\def\be{\begin{equation}}
\def\ee{\end{equation}}
\def\bea{\begin{eqnarray}}
\def\eea{\end{eqnarray}}

\usepackage{subcaption}
\usepackage{amsmath}
\usepackage{amssymb}
\usepackage{slashed}

\begin{document}
\vspace*{4cm}
\title{Parameter space for testable leptogenesis \footnote{Contribution to the 2023 Electroweak session of the 57$^{\rm{th}}$ Rencontres de Moriond based on the results from Ref. \cite{Drewes:2021nqr}.}}

\author{Yannis Georis}

\address{Centre for Cosmology, Particle Physics and Phenomenology, Université catholique de Louvain, Louvain-la-Neuve B-1348, Belgium}

\maketitle\abstracts{
Extending the Standard Model with right-handed neutrinos provides a minimal explanation for both light neutrino masses (through the type-I seesaw mechanism) and the baryon asymmetry of our universe (through leptogenesis). We map here for the first time the range of heavy neutrino mixing angle consistent with both neutrino masses and leptogenesis in a scenario with 3 generations of right-handed neutrinos with Majorana masses between 50 MeV and 70 TeV. Due to  the presence of a third degree of freedom that remains much more feebly coupled to the Standard Model thermal bath, we observe that the parameter space is much larger compared to the minimal scenario with 2 generations. This greatly enhances the testability prospects for low-scale leptogenesis and, in the most optimistic scenario, would allow experimentalists not only to discover right-handed neutrinos but also to perform consistency checks of the model.}

\section{Introduction}
While the Standard Model (SM) has been extremely successful over the past few decades, there remains some observations that it cannot account for. Among those are the light neutrino masses but also the absence in the present universe of anti-baryons in sizeable quantity. The latter asymmetry is usually parameterised by the baryon-to-entropy ratio \cite{Planck:2018vyg} at the CMB epoch
\begin{equation}
    Y_{B,\rm{obs}} = \frac{n_B - n_{\bar{B}}}{s} =(8.66 \pm 0.05)\cdot 10^{-11}.
\end{equation}
While the Standard Model can qualitatively explain the presence of a baryon asymmetry (BAU), it does not provide enough CP-violation \cite{Gavela:1993ts,Gavela:1994dt} and deviation from thermal equilibrium \cite{Shaposhnikov:1987tw} to explain it quantitatively. As we will see in the next sections, in addition to explaining light neutrino masses, right-handed neutrinos (RHN) can also provide the necessary additional sources of CP-violation and deviation from thermal equilibrium.

\section{Low-scale seesaw and leptogenesis}
Given that right-handed neutrinos are singlets under all gauge interactions, the most general renormalisable Lagrangian that can be written using only SM fields and right-handed neutrinos $\nu_R$ is given by the \textit{type-I seesaw} Lagrangian 
\begin{equation}
    \mathcal{L}_I  = \mathcal{L}_{\rm{SM}} + \bar{\nu}_R i\slashed{\partial} \nu_R - F_{\alpha i} \bar{l}_{L\alpha} \Tilde{\Phi} \nu_{Ri} - \frac{1}{2} \bar{\nu}_{Ri}^c M_{Mij} \nu_{Rj} + \rm{h.c.}
    \label{eq:typeIseesawLagrangian}
\end{equation}
$M_M$ is the Majorana mass matrix for right-handed neutrinos while $F$ is the matrix of Yukawa couplings between the left-handed doublets $l_L$ and the right-handed neutrinos and $\Tilde{\Phi} = i\sigma_2 \Phi^*$, $\Phi$ being the Higgs field. Below the electroweak symmetry breaking, $\Phi$ gets a non-zero vacuum expectation value $v$ and a diagonalisation of the mass term leads to the presence of \textit{light} and \textit{heavy} neutrinos, $\nu$ and $N$, whose mass is given at tree level by
\begin{equation}
    m_\nu = -v^2 F M_M^{-1} F^t \mbox{ and } M_N = M_M + \frac{1}{2}\left(\theta^\dagger\theta  M_M+ M_M^t \theta^t \theta^* \right),
    \label{eq:typeIseesawrelation}
\end{equation}
where $\theta$ is the heavy neutrino mixing angle defined as $vF M_M^{-1}$. From Eq. \eqref{eq:typeIseesawrelation}, we can deduce that one needs at least $n=2$ generations of right-handed neutrinos to explain the two observed non-zero light neutrino masses. Many experiments, both at colliders or fixed target, are currently searching for such new particles. One usually expresses experimental sensitivities in terms of the heavy neutrino interaction strength with the Standard Model flavour $\alpha$, $U_\alpha^2$, or its total interaction strength $U^2$ defined as
\begin{equation}
    U_\alpha^2 = \sum_{i=1}^3 |\theta_{\alpha i}|^2 = \left(\theta \theta^\dagger\right)_{\alpha\alpha} \mbox{ and } U^2 = \sum_{\alpha=e,\mu,\tau} U_\alpha^2.
\end{equation}
On the cosmology frontier, right-handed neutrinos could hold the key for matter-genesis. In the minimal \textit{vanilla} scenario of thermal leptogenesis \cite{Fukugita:1986hr}, C- and CP-violation are provided by the decays of heavy neutrinos during their freeze-out. However, although appealing, this scenario can only work for large heavy neutrino masses \cite{Davidson:2002qv} $M \gtrsim 10^9$ GeV, untestable at present-day experiments. Motivated by testability prospects, models with relatively low RHN masses have been developed over the past 30 years for which tiny mass splittings enhance the CP-violation (\textit{resonant leptogenesis} \cite{Pilaftsis:2003gt}) or where the baryon asymmetry is instead produced during the RHN freeze-in by CP-violating RHN oscillations (\textit{leptogenesis from neutrino oscillations} \cite{Akhmedov:1998qx}). While initially considered as two distinct mechanisms, a unified description of these two scenarios was recently presented in Ref. \cite{Klaric:2020phc}, see Eq. (1) in the latter reference for the set of equations describing the time evolution of RHN densities and SM lepton asymmetries.

\section{Viable parameter space for the 3 generations scenario}

Since RHNs are gauge singlets, their number of generations is in principle not restricted. While the minimal scenario $n=2$ has been already extensively studied \cite{Klaric:2020phc,Asaka:2005pn,Antusch:2017pkq,Hernandez:2016kel}, a comprehensive study of the parameter space for the case with three heavy neutrinos had yet to be performed which was the main focus of this work. Albeit less minimal, such a scenario can be motivated in the context of flavour symmetries or of SM gauge extensions, see \textit{e.g.} the Left-Right Symmetric Model, where anomaly cancellation requires $n=3$. We solved the RHN evolution equations for two types of initial conditions: 1) Vanishing initial conditions where heavy neutrinos are initially absent from the thermal bath and are only produced through their Yukawa couplings to SM neutrinos 2) Thermal initial conditions where heavy neutrinos are in thermal equilibrium at early time. The latter choice is well motivated in the context of SM gauge extensions. Indeed, while we remain here agnostic about a possible UV completion of the Lagrangian \eqref{eq:typeIseesawLagrangian}, it is conceivable that right-handed neutrinos would have at high temperature gauge interactions that would bring them in thermal equilibrium before the typical temperature scale of leptogenesis. Solving the RHN evolution equations, displayed in \textit{e.g.} Ref. \cite{Klaric:2020phc}, for RHN masses \footnote{Since we search to identify the largest and smallest $U^2$, we take the RHN mass splittings to be small $\frac{\Delta M}{\bar{M}} < 0.1$.} between 50 MeV and 70 TeV, we mapped the experimentally accessible parameter space consistent with neutrino masses and leptogenesis. Our results are shown in Fig. \ref{fig:radish}, our main findings being the following:

\begin{figure}[!t]
\begin{subfigure}{0.5\textwidth}
\centerline{\includegraphics[width=\linewidth]{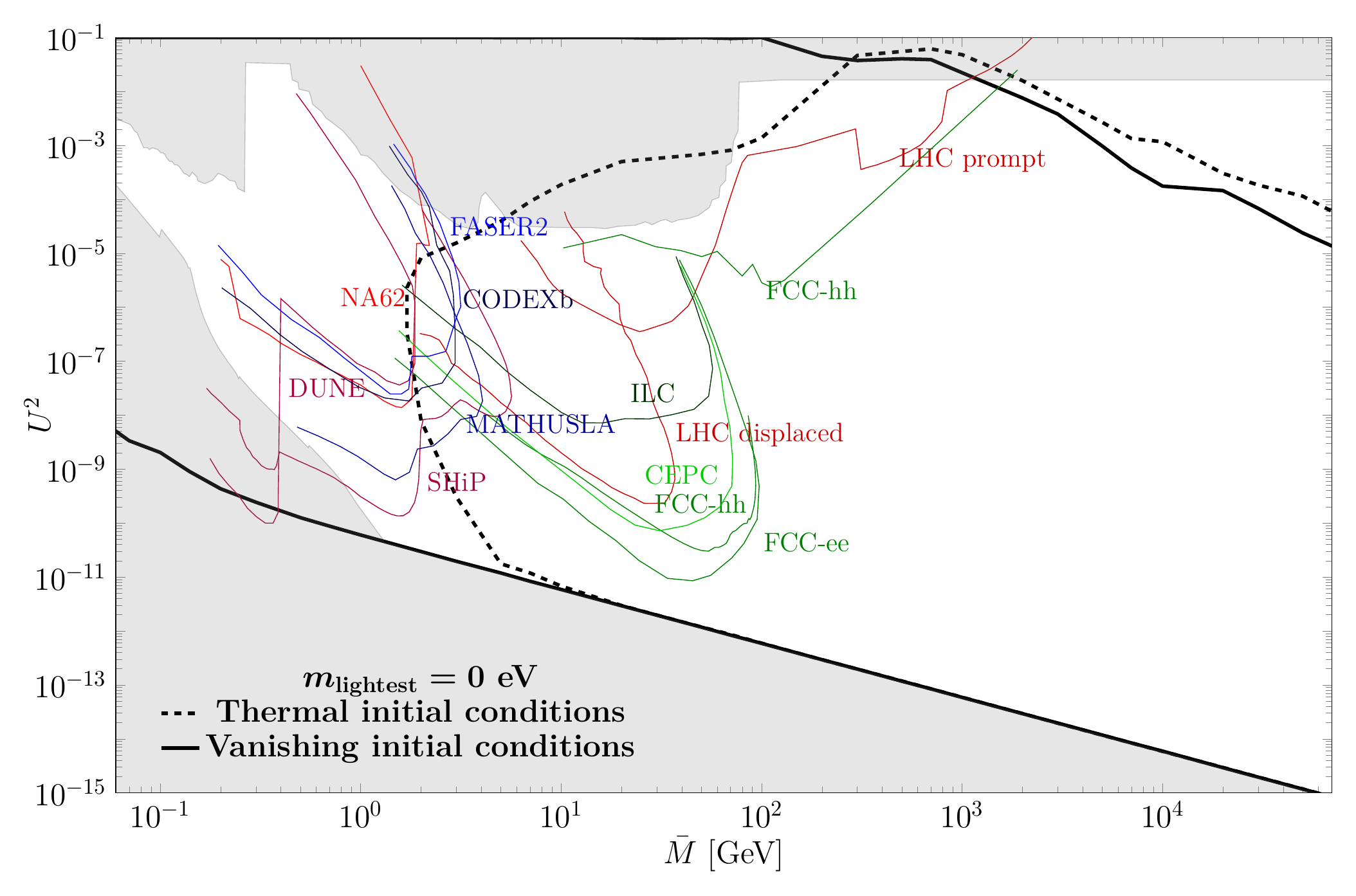}}
\end{subfigure}
\begin{subfigure}{0.5\textwidth}
\centerline{\includegraphics[width=\linewidth]{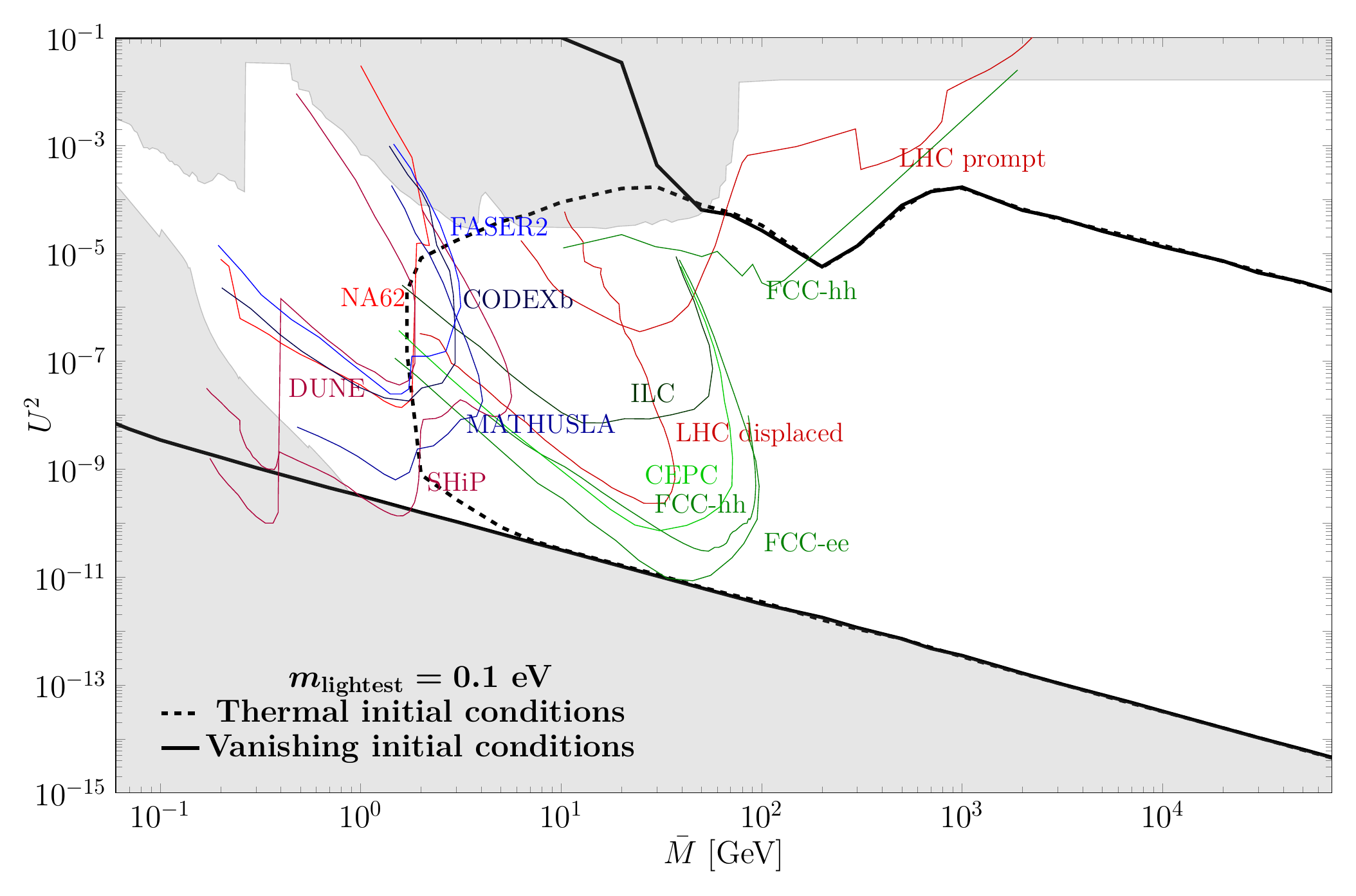}}
\end{subfigure}
\caption[]{We here show the viable parameter space in the $\bar{M}-U^2$ plane. Both light neutrino masses and the BAU for vanishing (thermal) initial conditions are reproduced within the continuous (dotted) black lines. The left (right) panel displays the results for a lightest neutrino mass $m_{\rm{lightest}} = 0$ eV ($=0.1$ eV). Experimental lines are displayed for a RHN coupling to muon only. Figures taken from Ref. \cite{Drewes:2021nqr}.}
\label{fig:radish}
\end{figure}
\vspace{-.24cm}
\begin{enumerate}
    \item The range of heavy neutrino mixing angle $U^2$ for which leptogenesis is viable is larger by several orders of magnitude compared to the results for $n=2$, compare with \textit{e.g.} Fig. 1 in Ref. \cite{Klaric:2020phc}. Many experiments, already running (\textit{e.g.} NA62, CMS, Atlas, ...) or planned to happen in the near-future (\textit{e.g.} HL-LHC, DUNE, ...), are expected to probe large part of this parameter space in the low mass regime. In the most optimistic scenario where the heavy neutrino mixing angle would lie just below the experimental limit, one could potentially observe thousands of RHNs at the HL-LHC such that one could \textit{e.g.} measure the branching ratio of RHN to each SM flavour with enough accuracy \cite{Drewes:2016jae} to perform a consistency check of RHN as sources for the light neutrino masses and the BAU. 
    \vspace{-.24cm}
    \item For thermal initial conditions, leptogenesis works for heavy neutrino masses much below the electroweak scale, as low as $1.7$ GeV.
    \vspace{-.24cm}
    \item The range of mixing angle consistent with leptogenesis depends on the value of $m_{\rm{lightest}}$. For large $m_{\rm{lightest}}$, the parameter space is slightly reduced albeit large mixing angles remain allowed.
    \vspace{-.24cm}
    \item We observe a dip in the parameter space at around $\bar{M}\sim 100$ GeV for $m_{\rm{lightest}} = 0.1$ eV where there is a shift between a regime where most of the BAU is produced during freeze-in to a regime where most of the BAU is produced during freeze-out. 
\end{enumerate}
These features can be explained by realising that the third (additional) degree of freedom remains much  more decoupled compared to the first two as it does not have to contribute sizeably to the light neutrino masses. In the context of freeze-in leptogenesis, this in particular implies that it can remain out of equilibrium for a very long period of time and only equilibrate at the sphaleron freeze-out temperature. The BAU would then also be produced right before the sphaleron freeze-out and avoid washout. An illustration of this effect is displayed in Fig. \ref{fig:LateBAUprodfreeze-in}.

\begin{figure}[!t]
    \centering
    \includegraphics[width=.49\textwidth]{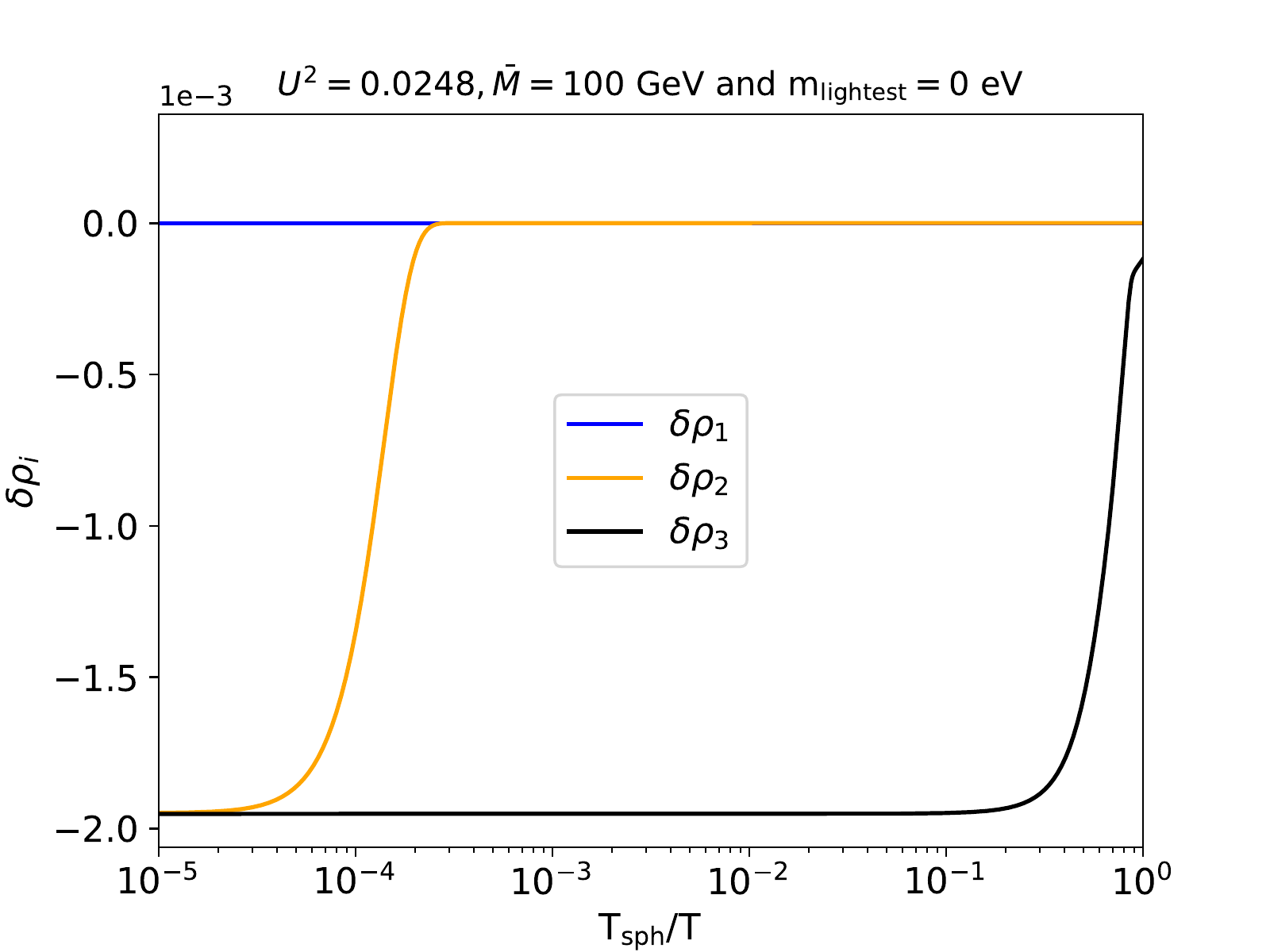}
    \includegraphics[width=.49\textwidth]{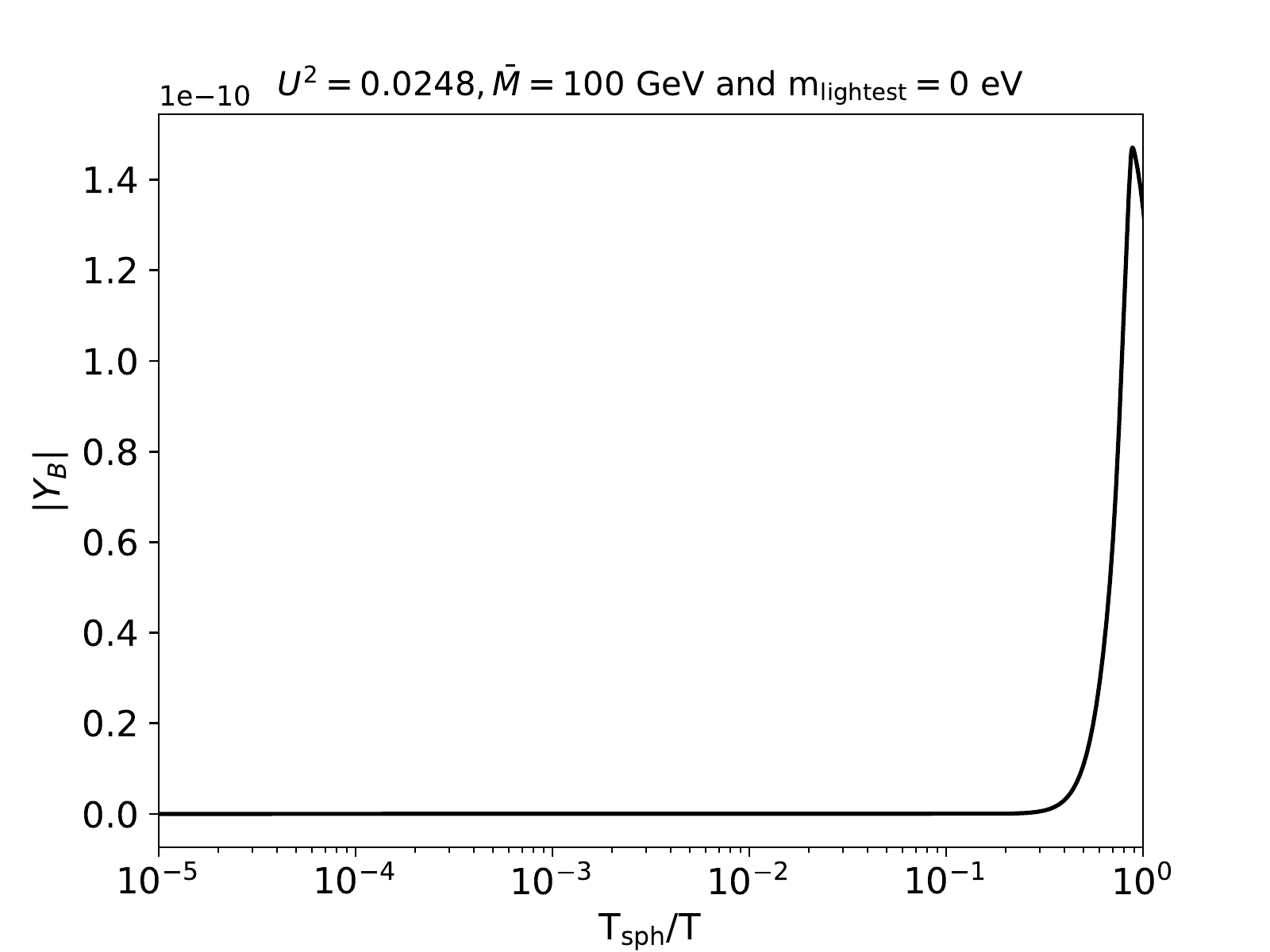}
    \caption{Illustration of the late BAU production mechanism for a benchmark scenario with a heavy neutrino mass $\bar{M} = 100$ GeV, $U^2 \simeq 0.0248 $ and for a massless lightest neutrino $m_{\rm{lightest}} = 0$ eV, see text for more details. The left and right panels display the evolution of RHN densities $\delta \rho_i$ and BAU $Y_B$ respectively.}
    \label{fig:LateBAUprodfreeze-in}
\end{figure}

In the context of freeze-out leptogenesis, having a third decoupled heavy neutrino enhances the deviation from thermal equilibrium and, therefore, the BAU production remains efficient even for relativistic RHNs. Finally, given that Yukawa couplings are proportional to the light neutrino mass spectrum, the coupling of the third heavy neutrino is directly proportional to the lightest neutrino mass $m_{\rm{lightest}}$. The third heavy neutrino is therefore more strongly coupled for $m_{\rm{lightest}}=0.1$ eV and this suppresses the BAU production at large mixing angles, explaining observation n°3.

\section*{Acknowledgments} 
YG acknowledges the support of the French Community of Belgium through the FRIA grant No. 1.E.063.22F. Computational resources have been provided by the supercomputing facilities of the Université catholique de Louvain (CISM/UCL) and the Consortium des Équipements de Calcul Intensif en Fédération Wallonie Bruxelles (CÉCI) funded by the Fond de la Recherche Scientifique de Belgique (F.R.S.-FNRS) under convention 2.5020.11 and by the Walloon Region.

\vspace{-.28cm}
\section*{References}

\bibliography{biblio}

\end{document}